\documentclass[aps,reprint,twocolumn,a4paper,superscriptaddress]{revtex4-1}

\usepackage{graphicx}
\usepackage{dcolumn}
\usepackage{bm}
\usepackage{setspace}
\usepackage{amsmath}

\begin{document}

\singlespacing

\title{Terahertz Kerr effect in a methylammonium lead bromide perovskite crystal}

\author{A. A. Melnikov}
\email{melnikov@isan.troitsk.ru}
\affiliation {Institute for Spectroscopy RAS, Fizicheskaya 5, Troitsk, Moscow, 108840 Russia}
\author{V. E. Anikeeva}
\affiliation {Institute for Spectroscopy RAS, Fizicheskaya 5, Troitsk, Moscow, 108840 Russia}
\affiliation {HSE University, Myasnitskaya ul. 20, Moscow 101000, Russia}
\author{O. I. Semenova}
\affiliation{A.V. Rzhanov Institute of Semiconductor Physics RAS, Novosibirsk 630090, Russia}
\author{S. V. Chekalin}
\affiliation {Institute for Spectroscopy RAS, Fizicheskaya 5, Troitsk, Moscow, 108840 Russia}

\begin{abstract}

We have observed short-lived optical birefringence in a CH$_3$NH$_3$PbBr$_3$ single crystal induced by a powerful nearly single-cycle terahertz pulse. Apart from the instantaneous contribution that follows the profile of the squared pump electric field, the recorded anisotropy signal contains an exponential component decaying in $\sim$ 350 fs, underdamped oscillations at the frequency of $\sim$ 0.16 THz and an intermediate picosecond relaxation process with a Gaussian tail. We associate these three non-trivial features with, respectively, Kerr effect in the inorganic lattice, terahertz-induced transient alignment of CH$_3$NH$_3^+$ cations, and their coherent rotation excited by the terahertz pulse in a Raman process.

\end{abstract}

\maketitle

\section{Introduction}

Lead-halide perovskites (LHPs) have the chemical formula ABX$_3$ common to all perovskite compounds. Inorganic atoms B = Pb and X = Cl, Br or I form a framework of BX$_6$ octahedra, while A-cations are located at the centers of cages formed by the octahedra \cite{Baikie, Brenner}. The key distinction of hybrid LHPs is that their A-cations are organic molecules, such as CH$_3$NH$_3^+$. In contrast to single atom cations of fully inorganic LHPs CH$_3$NH$_3^+$ has rotational degrees of freedom inside the cage and possesses a permanent dipole moment. Via their dipole moments organic cations can interact with the inorganic lattice, with neighboring cations, and with free charge carriers introduced e.g. via light absorption or electrical doping. Due to coexistence of the polarizable inorganic framework with the sublattice of polar rotating cations a hybrid LHP crystal represents a rather exotic system, which is of great interest from the fundamental point of view. In order to describe and interpret the resulting complex physics, the concepts that are rather unusual for semiconductors are often employed by drawing analogies with polar liquids \cite{Onoda, Frost, Zhu}, glasses \cite{Fabini, Simenas}, and plastic crystals \cite{Even}. 

Interactions between the cations and the charge carriers cause favorable transport properties of hybrid LHPs that triggered their fast rise as unique materials for photovoltaics \cite{Jena}. Carriers in the conduction and valence bands are believed to be stabilized as large polarons formed through the coupling with the ensemble of polar cations \cite{Zhu2}. Thereby their interaction with charged defects is screened promoting moderate carrier mobility and large diffusion lengths in inherently defect rich hybrid LHPs \cite{Stranks}. The electron-hole recombination is rather slow in hybrid LHPs \cite{Brenner}, which is also interpreted as the effect of cation dynamics. A possible reason is that the local orientational ordering of cations distorts the inorganic lattice, which makes the bandgap indirect, decreasing carrier recombination rate \cite{Motta}. The concomitant dipole ordering creates macroscopic electric fields that induce separation of charge carriers.

Since rotational dynamics of organic cations is of crucial importance for the physics of hybrid LHPs, it has been studied by various theoretical and experimental methods. Among them are calorimetry \cite{Fabini}, neutron scattering \cite{Swainson, Leguy, Li}, and various types of spectroscopy (NMR \cite{Wasylishen, Qiao}, microwave \cite{Poglitsch}, infrared \cite{Onoda2, Boldyrev}, Raman \cite{Guo}, and ultrafast \cite{Zhu, Bakulin}). As a result, a general picture has emerged, according to which at low temperatures the rotational motion is limited to librations around specific directions inside the cage with occasional reorientational jumps, while at higher temperatures in the cubic phase it can be treated as quasi-free rotation \cite{Leguy}. The relative timescales of the motion were also roughly determined allowing to distinguish sub-picosecond wobbling and slower reorientation \cite{Bakulin}. However, the characteristic values obtained for the same LHP compounds can vary by almost two orders of magnitude. There is still no clear unified picture of organic cation dynamics and its relation to other fundamental processes in LHPs remains not fully understood.

One of the ultrafast spectroscopic methods that were used to study cation dynamics in LHPs is the observation of the transient optical Kerr effect. This is a well-established tool for the investigation of molecular orientational dynamics \cite{Righini, McMorrow}. Ultrashort laser pulses are used to exert a torque on molecules in a medium under study via their hyperpolarizability. The result of this action is a small average molecular alignment that reveals itself as transient optical birefringence of the sample, provided the molecules are non-spherically symmetric. This birefringence can be measured with a temporally delayed probe laser pulse in the standard polarization-resolved pump-probe experiment. The recorded decay of the Kerr signal contains valuable information on the orientational relaxation, since it is approximately proportional to the correlation function of collective polarizability, while the latter is defined mainly by single-molecule and pair orientational correlation functions \cite{Kohler}. Femtosecond optical Kerr effect in both hybrid and inorganic LHP crystals was studied in Refs. \cite{Zhu, Miyata} where an evidence of the liquid-like response was obtained in the form of a picosecond non-exponential relaxation. However, in a recent publication \cite{Maehrlein} it was claimed to be an experimental artifact due to propagation effects experienced by femtosecond near-resonant pulses. 

In the present work we study the optical Kerr effect induced in a CH$_3$NH$_3$PbBr$_3$ single crystal by a powerful terahertz pulse and find signatures of collective rotational dynamics of methylammonium cations. Unlike femtosecond laser pulses, single-cycle terahertz pulses act directly on the permanent electric dipole of a polar molecule thereby exerting a torque on the latter. As a result, both the alignment and the orientation of the molecules occur, though only the former is accessible if the birefringence of the sample is probed. One more key difference is that the characteristic timescale of variation of the electric field of the terahertz pulse is $\sim$ 1 ps compared to $\sim$ 1 fs for femtosecond laser pulses. In combination with the high strength of the electric field this leads to a relatively high amplitude of induced molecular rotation during the pulse. Therefore, the detected Kerr signal can be considerably stronger in the case of terahertz excitation \cite{Sajadi}. 

\section{Experimental details}

\begin{figure}
\centering
\includegraphics{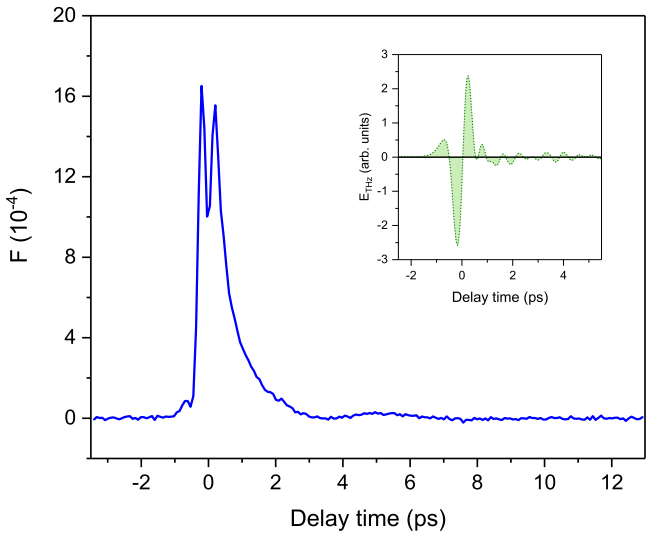}
\caption{\label{fig1} Transient optical anisotropy of the sample induced by the pump terahertz pulse. The inset shows a typical time dependence of the electric field strength of the pulse.}
\end{figure}

In our experiments we studied a 1.8 mm thick CH$_3$NH$_3$PbBr$_3$ single crystal grown from solution using the technique described in Refs. \cite{Semenova} and \cite{Yudanova}. The original method was developed for CH$_3$NH$_3$PbI$_3$, and the following modifications were made in order to adapt it for bromide crystal growth. CH$_3$NH$_3$Br precursor was synthesized by adding dropwise 28 ml of aqueous (40\%) methylamine solution to 44 ml of aqueous (48\%) HBr solution without preliminary H$_3$PO$_2$ stabilization. The synthesis proceeded for 2 hours during the reaction with continuous stirring at 0 $^\circ$C. A precipitate of yellowish color obtained by evaporation of the reaction mixture at 60--70 $^\circ$C was then washed with acetone to produce snow-white crystals. PbBr$_2$ was synthesized dissolving 28 g of crystalline lead acetate hydrate in 78.5 ml of an aqueous (48\%) HBr solution. At the final stage 18.45 g of CH$_3$NH$_3$Br were added to the flask with the synthesized PbBr$_2$, resulting in an orange precipitate of perovskite CH$_3$NH$_3$PbBr$_3$. CH$_3$NH$_3$PbBr$_3$ single crystals were grown from the obtained powder in the same way as CH$_3$NH$_3$PbI$_3$ single crystals. The crystal that was investigated in our experiments had a polyhedral shape somewhat similar to a rhombic dodecahedron. Therefore we tentatively identified the studied surface as (110).

Though CH$_3$NH$_3$PbBr$_3$ crystals are relatively stable under ambient conditions, one can expect certain modification of the crystal surface and adjacent layers as a result of regular exposure to air during long repeated measurements. Several months after the experiments reported in the present work were performed, we have repeated measurements with fresh samples and observed generally similar transient signals with a certain difference that was caused naturally by a variation of the pump terahertz waveform.

In order to detect the Kerr response of the crystal we used the standard pump-probe layout, in which a pump terahertz pulse excites the sample, while a temporally delayed probe pulse is used to measure the induced birefringence. All experiments were performed at room temperature and in ambient air. Nearly single-cycle terahertz pulses with a duration of $\sim$ 1 ps and peak electric field strength of up to $\sim$ 500 kV/cm were generated using the technique of optical rectification of femtosecond pulses with tilted fronts. The details of the method can be found elsewhere \cite{Stepanov, Melnikov}. The direction of the pump terahertz pulse polarization vector relative to the sample crystal axes was arbitrary. Weak femtosecond laser pulses of $\sim$ 50 fs duration at 800 nm were used for probing. Before passing through the sample the probe pulse was polarized at 45$^{\circ}$ relative to the vertical polarization direction of the terahertz electric field. After the sample the polarization of the probe pulse was rotated by $\sim$ 20$^{\circ}$ with a half-wave plate. Then the probe beam was transmitted through a Wollaston prism in order to split it into vertically and horizontally polarized beams. The latter were guided to the amplified photodiodes, peak amplitudes of the signals from which were proportional to the intensities of the orthogonally polarized components $I_x$ and $I_y$. The result of each individual measurement was a dependence of the quantity $F=(1-I_x/I_y)_{\mathrm{pump\;on}}\div(1-I_x/I_y)_{\mathrm{pump\;off}}$ on the delay time between the terahertz and the probe pulses. If $n_x$ and $n_y$ are the refractive indices of the sample experienced by horizontally and vertically polarized probe pulses respectively, and $\Delta n = n_x - n_y$, then $F \propto \Delta n$ in the limit of low $\Delta n$. We note here that without the additional rotation of polarization after the sample by the half-waveplate the detected signal would be close to zero, since  the appearance of a small $\Delta n$ induces ellipticity of the probe beam, while the major axis of that ellipse remains directed at 45$^{\circ}$ to the pump electric field.

\begin{figure}
\centering
\includegraphics{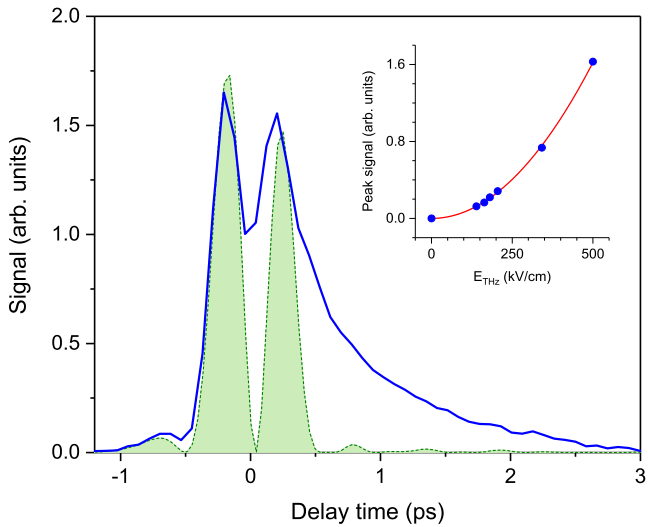}
\caption{\label{fig2} Comparison of the initial evolution of the optical anisotropy signal (solid line) with the squared electric field of the pump terahertz pulse (dashed line with filled area below it). The inset demonstrates the dependence of the peak amplitude of the $F(t)$ signal (circles) on the peak electric field and its parabolic fit by the $y=ax^2$ function (solid line).}
\end{figure}

\section{Results and discussion}

The optical anisotropy signal $F(t)$ detected using pump terahertz pulses with the highest available peak electric field strength of $\sim$ 500 kV/cm is shown in Fig. 1. The inset illustrates time dependence of the pump electric field $E_\mathrm{THz}(t)$. This temporal profile was measured via electro-optic detection of an attenuated terahertz pulse in a ZnTe crystal. The attenuation was necessary, since such strong terahertz fields cannot be measured correctly by this method because of pronounced nonlinear effects in ZnTe. It was performed by a 1 mm thick fused silica plate covered by a thin metal film (total optical density of $\sim$ 4 in the visible range). The plate caused a certain effective lengthening of the terahertz pulse due to high-frequency absorption and dispersion of fused silica in the terahertz range. Though this effect was visible in the experimental data (see below), it was rather small, so that it was possible to treat the measured waveform as a good approximation of the real pump terahertz pulse. Weak electric field oscillations after $\sim$ 1 ps were caused by the absorption of the terahertz radiation in air and were neglected, since the studied effect was quadratic in electric field. 

Specific relaxation components that comprise the $F(t)$ signal are relatively well separated in the time domain, which makes them easier to disentangle. It is possible to select three characteristic time windows. In the first one, roughly from $-1.0$ ps to 0.5 ps, the probe pulse overlaps with the pump terahertz pulse and the induced anisotropy is determined mainly by the electric field of the latter. The second window spans from $\sim$ 0.5 ps to $\sim$ 3 ps and contains a monotonically decaying component of the signal. The third window from $\sim$ 3 ps to $\sim$ 12 ps is occupied by weak low-frequency underdamped oscillations. Below we extract parameters of these temporal components and relate them to specific physical processes initiated in the CH$_3$NH$_3$PbBr$_3$ crystal by the pump terahertz pulse.

In Figure 2 we compare the starting part of the optical anisotropy signal $F(t)$ with the squared electric field $E_\mathrm{THz}^2(t)$. It can be seen that the positions of first three peaks of  $E_\mathrm{THz}^2(t)$ are slightly wider spaced than those of $F(t)$. This temporal spreading was introduced by the electric field measurement procedure and was a result of the attenuation of terahertz radiation, as discussed above. Here this effect can be neglected and the corresponding peaks of $E_\mathrm{THz}^2(t)$ and $F(t)$ treated as synchronous. As follows from Fig. 2, the relative amplitudes of the peaks are very similar for both transients and $F(t)$ almost ``follows'' the $E_\mathrm{THz}^2(t)$ waveform in the region of the pump and probe pulse temporal overlap. Therefore, we associate the initial part of the optical anisotropy signal from $-1.0$ ps to 0.5 ps with the instantaneous Kerr effect of electronic origin. It is hardly possible to ascribe it exclusively to the inorganic lattice of the CH$_3$NH$_3$PbBr$_3$ crystal or to its sublattice of organic cations since both subsystems can in principle contribute to the Kerr signal of this type. 

We have additionally measured the dependence of the peak $F(t)$ value on the peak strength of the pump electric field. The result is presented in the inset to Fig. 2. The obtained set of experimental points can be rather well fitted by a parabola $y=ax^2$ confirming the quadratic dependence. The intermediate picosecond relaxation in the time window from $\sim$ 0.5 ps to $\sim$ 3 ps also demonstrated quadratic variation with the electric field strength. However it was practically impossible in our experiments to measure properly the field dependence of the amplitude of oscillations because of its very small value. 

The observed similarity of the initial part of the $F(t)$ signal and the squared electric field waveform  $E_\mathrm{THz}^2(t)$ implies that in our experiment the temporal resolution is maintained despite possible propagation effects. Indeed, it is reasonable to expect that group refractive indices for the pump terahertz and the 800 nm probe pulses will differ considerably. It is particularly true for solids with a rich spectrum of dipole-active phonons, like LHPs. This difference implies a corresponding large difference of group velocities of the pulses. As a result, in a sufficiently thick transparent sample the probe pulse will ``scan'' the pump-induced response during propagation through the former and despite the fixed pump-probe delay time the signal from a wide time window will be integrated. In this case the temporal resolution decreases, while the signal becomes distorted. This effect can be approximately visualized by performing the convolution of a well-resolved signal with a ``pulse'' of a rectangular shape, the duration of which depends on the values of the group refractive indices. We have performed such calculations for the measured $E_\mathrm{THz}^2(t)$ profile and found that for $F(t)$ to have the observed specific shape in the interval $-1.0$ ps $< t <$ 0.5 ps, the total time lag between the pump and the probe pulses should not exceed $\delta \tau \sim$ 300 fs after traveling through the sample. This time can be estimated also as $\delta \tau \sim L (n_\mathrm{THz}-n^{gr}_\mathrm{pr})/c$, where $L$ is the sample thickness, $n_\mathrm{THz}$ and $n^{gr}_\mathrm{pr}$ are the refractive index at pump wavelength and the group refractive index at probe wavelength, respectively.

While $n_\mathrm{pr}$ can be readily found in literature \cite{Alias}, the available data on the optical properties of CH$_3$NH$_3$PbBr$_3$ in the terahertz range are scarce. Here we rely on the values of $n_\mathrm{THz}$ provided in Ref. \cite{Andrianov} for thin films of CH$_3$NH$_3$PbBr$_3$. We take $n_\mathrm{THz} = 8$ at 1 THz and $n^{gr}_\mathrm{pr} \approx n_\mathrm{pr} = 2$ at 800 nm. Using the value and the formula for $\delta \tau$ mentioned above we obtain as a rough estimate that the sample thickness should be not larger than 15 $\mu$m, which is much less than the actual thickness of the CH$_3$NH$_3$PbBr$_3$ crystal (1.8 mm). The values of extinction ($\kappa_\mathrm{THz} \approx 5$) and absorption ($\alpha \approx 200\; \mathrm{cm}^{-1}$) coefficients from Ref. \cite{Andrianov} and Ref. \cite{Maeng} imply that the absorption length for the radiation at 1 THz in CH$_3$NH$_3$PbBr$_3$ is $\sim$ 5 $\mu$m and $\sim$ 50 $\mu$m, respectively. Therefore, we can conclude that the observed signal originates from a sufficiently thin layer of the crystal (on the order of $\sim$ 10 $\mu$m) and the propagation effects can be neglected. 

\begin{figure}
\centering
\includegraphics{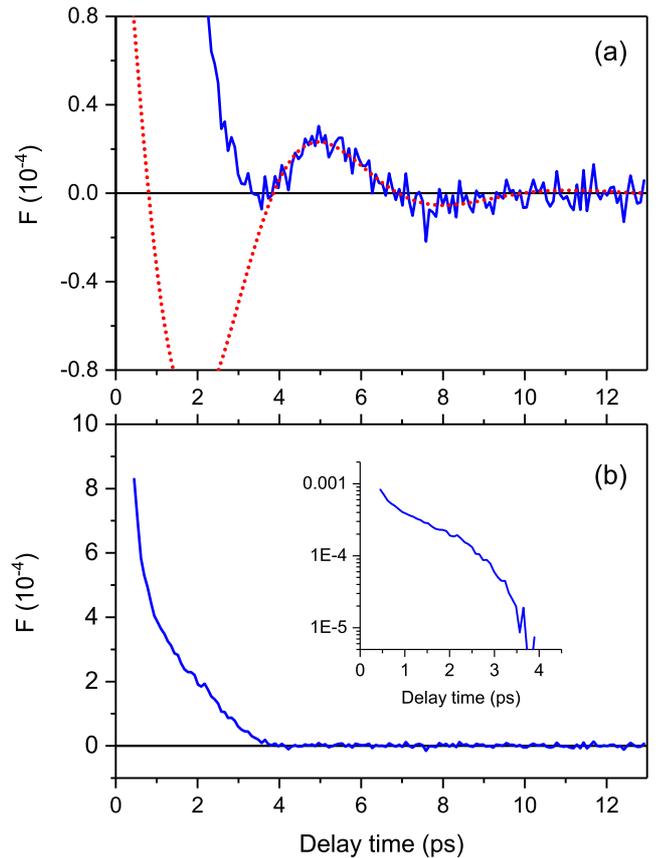}
\caption{\label{fig3} (a) -- Magnified view of the oscillatory part of the signal (solid line) and its fit (dotted line). (b) -- The residual monotonic component of the signal. The inset shows it in the logarithmic scale along the y-axis.}
\end{figure}

Next we proceed to the analysis of the oscillations that dominate the $F(t)$ signal  from $\sim$ 3 ps to $\sim$ 12 ps. The magnified view of this region is shown in Fig. 3(a). We have found that they can be well fitted by the function $A_1\exp(-t/\tau_{osc})\sin((2\pi/T)t+\phi)$, where the lifetime and the period of the oscillations are $\tau_{osc}=2.1\pm0.3$ ps and $T=6.1\pm0.2$ ps. The latter corresponds to a frequency of $164\pm5$ GHz ($5.5\pm0.2$ cm$^{-1}$ or $0.68\pm0.02$ meV). Such a low frequency can hardly be associated with the vibrations of inorganic framework of CH$_3$NH$_3$PbBr$_3$. Indeed, the lowest frequency phonons of LHP crystals that are due to collective dynamics of PbX$_6$ octahedra (tilting, rotation) or periodic translational motions of cations inside inorganic cages (rattling modes) have energies of several meVs and higher \cite{Ferreira, Zhao, Ponce, Song, Swainson, Atkins}. Probably the lowest energy of a phonon mode in a LHP crystal reported to date was measured for cooperative two-dimensional rotations of PbBr$_6$ octahedra in a fully inorganic CsPbBr$_3$ crystal ($\sim$1 meV) \cite{Atkins}. However, these are zone-boundary phonons, which cannot be accessed by optical probing. Besides, such low-frequency modes in hybrid LHPs are typically overdamped at room temperature and no long-lived coherence can be generated via their impulsive excitation.  

It is thus natural to suppose that the observed oscillations are caused by coherent rotational dynamics of organic CH$_3$NH$_3^+$ cations (here by “rotational dynamics” we mean rotation around the axis that is perpendicular to the C-N axis). This mechanism is essentially the impulsive stimulated rotational Raman scattering and implies that the pump terahertz pulse creates a coherent superposition of wave functions of rotational levels of CH$_3$NH$_3^+$ coupled via allowed Raman transitions. For the simplest case of linear molecules the radiation couples rotational levels, the angular momenta of which differ by $\Delta J=2$, which is a necessary condition for the observation of periodically modulated optical anisotropy of the excited molecular ensemble. Then the frequency of the observed oscillations of pump-induced birefringence is $\nu={\Delta E}/h = 2B(2J+3)$, where $J$ is the angular momentum of the lower-lying of the two rotational states, and $B$ is the rotational constant in cm$^{-1}$. Similar rules can be written for rigid symmetric top molecules, but for more complex rotors the information on exact positions of rotational levels should be obtained from experiments or theoretical calculations. Interestingly, in a recent study of terahertz Kerr effect in water coherent oscillations were observed in the optical anisotropy response of the gaseous phase at elevated temperatures \cite{Elgabarty}. The frequency of those oscillations was $\approx$ 37 cm$^{-1}$, which is close to the lowest frequency line in the rotational Raman spectrum of water vapor \cite{Nordstrom, Avila}. 

Unfortunately, the rotational Raman spectrum of the CH$_3$NH$_3^+$ molecular cation is not known. However, if we applied the simple linear molecule model, we could predict coherent oscillations of terahertz-induced birefringence rather similar to those observed in our experiment. Taking $B\approx0.7$ cm$^{-1}$ (with a reference to the somewhat similar CH$_3$NH$_2$ molecule \cite{Herzberg}) we obtain the following possible oscillation frequencies $\nu_J$ (in cm$^{-1}$): 4.2, 7, 9.8, 12.6, ... for $J=$0, 1, 2, 3, ... . However, two factors favor excitation of only a few lowest frequencies from this series. First, the terahertz waves at frequencies $\nu^\mathrm{THz}_i$ and $\nu^\mathrm{THz}_k$, which take part in the elementary Raman scattering process, necessarily belong to the spectrum of the pump pulse. Therefore, since $\nu_J = \nu^\mathrm{THz}_i - \nu^\mathrm{THz}_k$, and the amplitude of coherent rotations is proportional to the product of amplitudes of the waves, the efficiency of scattering will decrease fast for higher $\nu_J$. In addition, this effect will be further enhanced by the $(\nu^\mathrm{THz}_i -\nu_J)^4$ frequency dependence of the Raman cross section.

In the framework of this model the measured frequency of 5.5 cm$^{-1}$ turns out to be very close to the mean frequency of the $\nu_0$ and $\nu_1$ model Raman lines and, therefore the observed oscillations can be the result of beating of these harmonics. Due to the strong damping of oscillations it is possible in principle to achieve a decent reproduction of the observed oscillatory kinetics by constructing a linear combination of the $\nu_0$ and $\nu_1$ modes with a certain adjustments of their phases and even with an additional smaller contribution of the next two or three modes. Associating the coherent rotational motion with a limited set of impulsively excited rotational transitions, the frequencies of which depend only on the inertial properties of the CH$_3$NH$_3^+$ molecule, allows an assumption that the decay of observed oscillations reflects the population relaxation. Then the extracted decay time $\tau_{osc}=2.1\pm0.3$ ps can be ascribed to the damping of rotations of individual CH$_3$NH$_3^+$ cations caused by their interaction with host inorganic cages. 

There is of course a number of important potential corrections, which can affect this simplified model considerably. Among them are a more precise definition of the rotational constant, calculation of the rotational spectrum taking into account nonzero values of the angular momentum around the molecule axis and possible rotations of the CH$_3$ and NH$_3$ groups, as well as a proper treatment of the stimulated scattering of the terahertz pulse with a broad spectrum of frequencies in the vicinity of rotational transitions. However, these refinements complicate the problem considerably and require a separate study that is beyond the scope of the present work.

\begin{figure}
\centering
\includegraphics{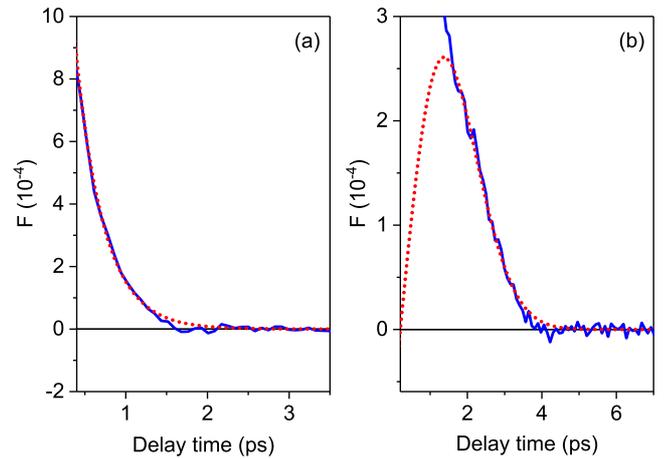}
\caption{\label{fig4} (a) -- The residual component of the monotonic signal after subtracting the function that fits the Gaussian-like tail (shown in (b)).}
\end{figure}

The part of the signal that remains after subtraction of the fitted oscillatory component is shown in Fig. 3(b) starting from the delay time of 0.5 ps. This monotonic relaxation has a clearly non-exponential character, which is emphasized in the inset to Fig. 3(b), where this curve is plotted in the logarithmic scale. Here it can be seen that the initial fast sub-picosecond decay is followed by a slower relaxation with a Gaussian-like tail.

This specific form of the Kerr effect decay curve (fast decrease with an adjacent ``shoulder'') is rather common for liquids in the case of excitation by a short femtosecond laser pulse \cite{Kohler, Quitevis, Chang}. The non-exponential part of the observed relaxation is often interpreted as laser-induced overdamped coherent librations of molecules in ``cages'' formed by their momentary neighbors. This process is modeled as a superposition of sine oscillators, the frequencies of which have a Gaussian distribution. However, characteristic periods of these librations are on the order of several hundreds of femtoseconds. The librations of organic cations in hybrid LHPs are believed to have similar typical characteristic times of $\sim$ 300 fs. This timescale is one order of magnitude shorter than the characteristic duration of the Gaussian ``shoulder'' in the Kerr response of CH$_3$NH$_3$PbBr$_3$ observed in our experiment.  Moreover, this type of motion is likely to be irrelevant for the room-temperature cubic phase of CH$_3$NH$_3$PbBr$_3$, when organic cation rotation is not restricted around a specific direction inside the inorganic cage. 

One more process that may appear in the ultrafast optical Kerr response of a molecular ensemble as a Gaussian-like decay is the relaxation of initially aligned molecules via inertial rotation. The characteristic timescale of this process can be estimated as $\sqrt{I/kT}$, where $I$ is the moment of inertia of the molecules \cite{Steele}. For CH$_3$NH$_3^+$ this value is $\sim$ 0.3 ps, which is also much faster than the observed picosecond decay. Moreover, the inertial decay of orientational correlations can be demonstrated in pure form only by free rotors, which is not the case of organic cations located in the inorganic cages. 

For molecules in liquids the timescale of several picoseconds is typical for the diffusive rotational motion. This regime supersedes the inertial relaxation for hindered rotors that interact with their environment and as a result experience frequent changes of the phase of rotation \cite{Fayer}. It should be noted that the diffusive decay of orientational correlations usually has exponential character and for liquids composed of symmetric top molecules it is even single-exponential. For certain more complex systems, such as ensembles of molecules embedded in a solid medium, the decay can be described by a stretched exponential function or by a power law \cite{Zanotti, Laage, Romero}. On the contrary, compressed exponential relaxation and its special case, Gaussian decay, are rather uncommon.

However, we have found that a rather good model of the monotonic part of the observed Kerr signal can be constructed if we assume that the transient alignment of the CH$_3$NH$_3^+$ molecules created by the pump terahertz pulse demonstrates Gaussian relaxation. In this case temporal evolution of the orientational correlation function of the organic cations is defined by the factor $e^{-(t/\tau)^2}$ assuming impulsive character of the terahertz excitation. Through the anisotropic polarizability of the molecules this behavior will be transferred to the correlation function of molecular polarizability $C(t)$ with the corresponding contribution to the Kerr signal in the form of $-dC(t)/dt\propto te^{-(t/\tau)^2}$ \cite{Kohler}. We have found that the latter function can serve as a rather good fit of the monotonic component of the detected signal starting from $\sim$1.5 ps, which is illustrated by Fig. 4(b). The fitting procedure provided the value $\tau=1.65\pm0.05$ ps. The residual shown in Fig. 4(a) represents a fast single exponential relaxation with a characteristic time of 350$\pm$10 fs, which can be associated with the Kerr response of the inorganic lattice of the CH$_3$NH$_3$PbBr$_3$ crystal in accordance with the interpretation proposed in Ref. \cite{Zhu}.

Gaussian decay observed here can be an evidence of cooperative effects during orientational relaxation of CH$_3$NH$_3^+$ cations. Here it is possible to draw an analogy with glassy materials, and in particular with metal glasses, for which a compressed exponential decay of the density correlation function was observed \cite{Trachenko}. It was suggested that relaxation of a local density fluctuation induces a long-range elastic deformation field, which stimulates structural relaxation at neighboring sites of the medium. In this way an ``avalanche'' of relaxation events is launched, resulting in the ensemble-averaged decay that is faster than exponential. In our case we study fluctuations of orientation of molecules and the timescales are orders of magnitude shorter. However, CH$_3$NH$_3^+$ cations rotating in the inorganic cages interact with each other via their permanent dipoles or via local lattice distortions that they create upon rotation. Thereby collective relaxational effects and Gaussian decay of terahertz-induced cation alignment may become possible. 

\section{Conclusion}

In conclusion, we observed the optical Kerr response of a CH$_3$NH$_3$PbBr$_3$ single crystal to an intense terahertz pulse. We decomposed the signal into several distinct decay components and suggested an assignment of each component to a certain relaxation process taking into account the hybrid structure of the perovskite. The initial fast sub-picosecond decay was interpreted as the Kerr effect in the inorganic lattice of the crystal. The slower relaxation on the timescale of several picoseconds was ascribed to the rotational dynamics of organic CH$_3$NH$_3^+$ cations. It comprises a Gaussian-like decay and weak underdamped low frequency oscillations. The former can be associated with diffusive relaxation of cation alignment created by the pump electric field, while the fact that it follows the compressed exponential law can be a signature of collective glass-like dynamics. Oscillations are due to coherent rotations of individual cations induced in the process of impulsive stimulated rotational Raman scattering of the terahertz pulse. The characteristic relaxation times determined in the framework of the suggested interpretation can be used to model interactions of organic cations with inorganic lattice and with each other. Observation of the strong-field terahertz Kerr effect in CH$_3$NH$_3$PbBr$_3$ and other hybrid lead halide perovskites at lower temperatures is also of great interest, since it implies studies of nonlinear dynamics of organic cations, the motion of which is initially restricted along certain directions inside inorganic cages. Since the applied method allows visualization of rotational dynamics in real time, it can be effectively combined with dynamic or quasistatic stimuli like excitation of electrons, application of external fields or heating in order to study interaction of organic cations with charge carriers and their dynamics at phase transitions.

\begin{acknowledgments}

The reported study was partially supported by the Academic Fund Program at the HSE University (grant number 21-04-016). 

\end{acknowledgments}


\begin{thebibliography}{99}

\bibitem{Baikie}
T. Baikie, Y. Fang, J. M. Kadro, M. Schreyer, F. Wei, S. G. Mhaisalkar, M. Graetzeld and T. J. White, Synthesis and crystal chemistry of the hybrid perovskite CH$_3$NH$_3$PbI$_3$ for solid-state sensitised solar cell applications, J. Mater. Chem. A  \textbf{1}, 5628 (2013).

\bibitem{Brenner}
T. M. Brenner, D. A. Egger, L. Kronik, G. Hodes and D. Cahen, Hybrid organic-inorganic perovskites: low-cost semiconductors with intriguing charge-transport properties, Nat. Rev. Mater. \textbf{1}, 15007 (2016).

\bibitem{Onoda}
N. Onoda-Yamamuro, T. Matsuo, H. Suga, Dielectric study of CH$_3$NH$_3$PbX$_3$, (X = Cl, Br, I), J. Phys. Chem. Solids \textbf{53}, 935-939 (1992).

\bibitem{Frost}
J. M. Frost and A. Walsh, What Is Moving in Hybrid Halide Perovskite Solar Cells?, Acc. Chem. Res. \textbf{49}, 528–535 (2016).

\bibitem{Zhu}
H. Zhu, K. Miyata, Y. Fu, J. Wang, P. P. Joshi, D. Niesner, K. W. Williams, S. Jin, X.-Y. Zhu, Screening in crystalline liquids protects energetic carriers in hybrid perovskites, Science \textbf{353}, 1409 (2016).

\bibitem{Fabini}
D. H. Fabini, T. Hogan, H. A. Evans, C. C. Stoumpos, M. G. Kanatzidis, and Ram Seshadri, Dielectric and thermodynamic signatures of low-temperature glassy dynamics in the hybrid perovskites CH$_3$NH$_3$PbI$_3$ and HC(NH$_2$)$_2$PbI$_3$, J. Phys. Chem. Lett. \textbf{7}, 376–381 (2016).

\bibitem{Simenas}
M. Simenas, S. Balciunas, J. N. Wilson, S. Svirskas, M. Kinka, A. Garbaras, V. Kalendra, A. Gagor, D. Szewczyk, A. Sieradzki, M. Maczka, V. Samulionis, A. Walsh, R. Grigalaitis, and J. Banys, Suppression of phase transitions and glass phase signatures in mixed cation halide perovskites, Nat. Commun. \textbf{11}, 5103 (2020).

\bibitem{Even}
J. Even, M. Carignano, and C. Katan, Molecular disorder and translation/rotation coupling in the plastic crystal phase of hybrid perovskites, Nanoscale \textbf{8}, 6222, (2016).

\bibitem{Jena}
A. K. Jena, A. Kulkarni, and T. Miyasaka, Halide perovskite photovoltaics: background, status, and future prospects, Chem. Rev. \textbf{119} 3036–3103 (2019).

\bibitem{Zhu2}
X.-Y. Zhu, and V. Podzorov, Charge Carriers in Hybrid Organic–Inorganic Lead Halide Perovskites Might Be Protected as Large Polarons, . Phys. Chem. Lett. \textbf{6}, 4758–4761 (2015).

\bibitem{Stranks}
S. D. Stranks, G. E. Eperon, G. Grancini, C. Menelaou, M. Alcocer, T. Leijtens, L. M. Herz, A. Petrozza, and H. J. Snaith, Electron-hole diffusion lengths exceeding 1 micrometer in an organometal trihalide perovskite absorber, Science \textbf{342}, 341 (2013).

\bibitem{Motta}
C. Motta, F. El-Mellouhi, S. Kais, N. Tabet, F. Alharbi, and S. Sanvito, Revealing the role of organic cations in hybrid halide perovskite CH$_3$NH$_3$PbI$_3$, Nat. Commun. \textbf{6}, 7026 (2015).

\bibitem{Swainson}
I. P. Swainson, C. Stock, S. F. Parker, L. Van Eijck, M. Russina, and J. W. Taylor, From soft harmonic phonons to fast relaxational dynamics in CH$_3$NH$_3$PbBr$_3$, Phys. Rev. B \textbf{92}, 100303(R) (2015).

\bibitem{Leguy}
A. M. A. Leguy, J. M. Frost, A. P. McMahon, V. G. Sakai, W. Kochelmann, C. Law, X. Li, F. Foglia, A. Walsh, B. C. O’Regan, J. Nelson, J. T. Cabral, P. R. F. Barnes, The dynamics of methylammonium ions in hybrid organic–inorganic perovskite solar cells, Nat. Commun. \textbf{6}, 7124 (2015).

\bibitem{Li}
B. Li, Y. Kawakita, Y. Liu, M. Wang, M. Matsuura, K. Shibata, S. Ohira-Kawamura, T. Yamada, S. Lin, K. Nakajima, S. F. Liu, Polar rotor scattering as atomic-level origin of low mobility and thermal conductivity of perovskite CH$_3$NH$_3$PbI$_3$, Nat. Commun. \textbf{8}, 16086 (2017).

\bibitem{Wasylishen}
R. E. Wasylishen, O. Knop, J. B. Macdonald, Cation rotation in methylammonium lead halides, Solid State Commun. \textbf{56}, 581-582 (1985).

\bibitem{Qiao} W.-C. Qiao, J. Wu, R. Zhang, W. Ou-Yang, X. Chen, G. Yang, Q. Chen, X. L. Wang, H. F. Wang, Y.-F. Yao, In situ NMR investigation of the photoresponse of perovskite crystal, Matter \textbf{3}, 2042 (2020).

\bibitem{Poglitsch}
A. Poglitsch and D. Weber, Dynamic disorder in methylammoniumtrihalogenoplumbates (II) observed by millimeter‐wave spectroscopy, J. Chem. Phys. \textbf{87}, 6373 (1987).

\bibitem{Onoda2}
N. Onoda-Yamamuro, T. Matsuo, H. Suga, Calorimetric and IR spectroscopic studies of phase transitions in methylammonium trihalogenoplumbates (II), J. Phys. Chem. Solids \textbf{51}, 1383-1395 (1990).

\bibitem{Boldyrev}
K. N. Boldyrev, V. E. Anikeeva, O. I. Semenova, and M. N. Popova, Infrared Spectra of the CH$_3$NH$_3$PbI$_3$ Hybrid Perovskite: Signatures of Phase Transitions and of Organic Cation Dynamics, J. Phys. Chem. C \textbf{124}, 23307–23316, (2020).

\bibitem{Guo}
Y. Guo, O. Yaffe, D. W. Paley, A. N. Beecher, T. D. Hull, G. Szpak, J. S. Owen, L. E. Brus, and M. A. Pimenta, Interplay between organic cations and inorganic framework and incommensurability in hybrid lead-halide perovskite CH$_3$NH$_3$PbBr$_3$, Phys. Rev. Materials \textbf{1}, 042401(R) (2017).

\bibitem{Bakulin}
A. A. Bakulin, O. Selig, H. J. Bakker, Y. L. A. Rezus, C. Muller, T. Glaser, R. Lovrincic, Z. Sun, Z. Chen, A. Walsh, J. M. Frost, and T. L. C. Jansen, Real-time observation of organic cation reorientation in methylammonium lead iodide perovskites, J. Phys. Chem. Lett. \textbf{6}, 3663-3669 (2015).

\bibitem{Righini}
R. Righini, Ultrafast Optical Kerr Effect in Liquids and Solids, Science \textbf{262}, 1386-1390 (1993).

\bibitem{McMorrow}
D. McMorrow, W. T. Lotshaw and G. A. Kenney-Wallace, Femtosecond optical Kerr studies on the origin of the nonlinear responses in simple liquids, IEEE J. Quantum Electron. \textbf{24}, 443-454, (1988).

\bibitem{Kohler}
B. Kohler and K. A. Nelson, Femtosecond impulsive stimulated light scattering from liquid carbon at high pressure: experiment and computer simulation, J. Phys. Chem. \textbf{96}, 6532 (1992).

\bibitem{Miyata}
K. Miyata, D. Meggiolaro, M. Tuan Trinh, P. P. Joshi, E. Mosconi, S. C. Jones, F. De Angelis, X.-Y. Zhu, Large polarons in lead halide perovskites, Sci. Adv. \textbf{3}, e1701217 (2017).

\bibitem{Maehrlein}
S. F. Maehrlein, P. P. Joshia, L. Huber, F. Wang, M. Cherasse, Y. Liu, D. M. Juraschek, E. Mosconi, D. Meggiolaro, F. De Angelis, and X.-Y. Zhu, Decoding ultrafast polarization responses in lead halide perovskites by the two-dimensional optical Kerr effect, Proc. Natl. Acad. Sci. USA \textbf{118}, e2022268118 (2021).

\bibitem{Sajadi}
M. Sajadi, M. Wolf, and T. Kampfrath, Transient birefringence of liquids induced by terahertz electric-field torque on permanent molecular dipoles, Nat. Commun. \textbf{8}, 14963 (2017).

\bibitem{Semenova}
O. I. Semenova, E. S. Yudanova, N. A. Yeryukov, Y. A. Zhivodkov, T. S. Shamirzaev, E. A. Maximovskiy, S. A. Gromilov, I. B. Troitskaia, Perovskite CH$_3$NH$_3$PbI$_3$ crystals and films. Synthesis and characterization, J. Cryst. Growth \textbf{462}, 45-49 (2017).

\bibitem{Yudanova}
E. S. Yudanova, T. A. Duda, O. E. Tereshchenko, O. I. Semenova, Properties of methylammonium lead iodide perovskite single crystals, J. Struct. Chem. \textbf{58} 1567-1572 (2017).

\bibitem{Stepanov}
A. G. Stepanov, J. Hebling, and J. Kuhl, Efficient generation of subpicosecond terahertz radiation by phase-matched optical rectification using ultrashort laser pulses with tilted pulse fronts, Appl. Phys. Lett. \textbf{83}, 3000 (2003).

\bibitem{Melnikov}
A. A. Melnikov, K. N. Boldyrev, Yu. G. Selivanov, V. P. Martovitskii, S. V. Chekalin, and E. A. Ryabov, Coherent phonons in a Bi$_2$Se$_3$ film generated by an intense single-cycle THz pulse, Phys. Rev. B \textbf{97}, 214304 (2018).

\bibitem{Alias}
M. S. Alias, I. Dursun, M. I. Saidaminov, E. M. Diallo, P. Mishra, T. K. Ng, O. M. Bakr, and B. S. Ooi, Optical constants of  CH$_3$NH$_3$PbBr$_3$ perovskite thin films measured by spectroscopic ellipsometry, Opt. Express \textbf{24}, 16586 (2016).

\bibitem{Andrianov}
A. V. Andrianov, A. N. Aleshin, A. O. Zakhar'in and L. B. Matyushkin, Terahertz Optical Characteristics of Organometallic Lead-Iodide (Bromide) Perovskites and Cesium Lead Halide Nanocrystals, \textit{ 2018 43rd International Conference on Infrared, Millimeter, and Terahertz Waves (IRMMW-THz)}, pp. 1-2 (2018).

\bibitem{Maeng} 
I. Maeng, S. Lee, H. Tanaka, J.-H. Yun, S. Wang, M. Nakamura, Y.-K. Kwon, and M.-C. Jung, Unique phonon modes of a CH$_3$NH$_3$PbBr$_3$ hybrid perovskite film without the influence of defect structures: an attempt toward a novel THz-based application, NPG Asia Mater. \textbf{12}, 53 (2020).

\bibitem{Ferreira}
A. C. Ferreira, S. Paofai, A. Létoublon, J. Ollivier, S. Raymond, B. Hehlen, B. Rufflé, S. Cordier, C. Katan, J. Even, and P. Bourges, Direct evidence of weakly dispersed and strongly anharmonic optical phonons in hybrid perovskites, Commun. Phys. \textbf{3}, 48 (2020). 

\bibitem{Zhao}
D. Zhao, J. M. Skelton, H. Hu, C. La-o-vorakiat, J.-X. Zhu, R. A. Marcus, M.-E. Michel-Beyerle, Y. M. Lam, A. Walsh, and E. E. M. Chia, Low-frequency optical phonon modes and carrier mobility in the halide perovskite CH$_3$NH$_3$PbBr$_3$ using terahertz time-domain spectroscopy, Appl. Phys. Lett. \textbf{111}, 201903 (2017).

\bibitem{Ponce}
S. Poncé, M. Schlipf, and F. Giustino, Origin of Low Carrier Mobilities in Halide Perovskites, ACS Energy Lett. \textbf{4}, 456 (2019).

\bibitem{Song}
F. Song, C. Qian, Y. Wang, F. Zhang, K. Peng, S. Wu, X. Xie, J. Yang, S. Sun, Y. Yu, J. Dang, S. Xiao, L. Yang, K. Jin, H. Zhong, X. Xu, Hot Polarons with Trapped Excitons and Octahedra‐Twist Phonons in CH$_3$NH$_3$PbBr$_3$ Hybrid Perovskite Nanowires, Laser Photonics Rev. \textbf{14}, 1900267 (2020).

\bibitem{Atkins}
T. Lanigan-Atkins, X. He, M. J. Krogstad, D. M. Pajerowski, D. L. Abernathy, Guangyong N. M. N. Xu, Zhijun Xu, D.-Y. Chung, M. G. Kanatzidis, S. Rosenkranz, R. Osborn, and O. Delaire, Two-dimensional overdamped fluctuations of the soft perovskite lattice in CsPbBr$_3$, Nat. Mater. \textbf{20}, 977 (2021).

\bibitem{Elgabarty}
H. Elgabarty, T. Kampfrath, D. J. Bonthuis, V. Balos, N. K. Kaliannan, P. Loche, R. R. Netz, M. Wolf, T. D. Kühne, M. Sajadi, Energy transfer within the hydrogen bonding network of water following resonant terahertz excitation. Sci. Adv. \textbf{6}, eaay7074 (2020).

\bibitem{Nordstrom}
E. Nordström, A. Bohlin, and P.-E. Bengtsson, Pure rotational Coherent anti-Stokes Raman spectroscopy of water vapor and its relevance for combustion diagnostics, J. Raman Spectrosc. \textbf{44}, 1322–1325 (2013).

\bibitem{Avila}
G. Avila, G. Tejeda, J. M. Fernandez, and S. Montero, The rotational Raman spectra and cross sections of H$_2$O, D$_2$O, and HDO, J. Mol. Spectrosc. \textbf{220}, 259–275 (2003).


\bibitem{Herzberg}
G. Herzberg, \textit{Molecular Spectra and Molecular Structure III. Electronic Spectra and Electronic Structure of Polyatomic Molecules}, Van Nostrand, New York (1966). 

\bibitem{Quitevis}
E. L. Quitevis and M. Neelakandan, Femtosecond optical Kerr effect studies of liquid methyl iodide, J. Phys. Chem. \textbf{100}, 10005 (1996).

\bibitem{Chang}
Y. J. Chang and Edward W. Castner, Jr., Intermolecular dynamics of substituted benzene and cyclohexane liquids, studied by femtosecond nonlinear-optical polarization spectroscopy, J. Phys. Chem. \textbf{100}, 3330 (1996).

\bibitem{Steele}
W. A. Steele, Molecular reorientation in liquids. II. Angular autocorrelation functions, J. Chem. Phys. \textbf{38}, 2411 (1963).

\bibitem{Fayer}
M. Tim, T. A. Jackson, P. A. Anfinrud in \textit{Ultrafast Infrared and Raman Spectroscopy}, edited by M. D. Fayer (Marcel Dekker, New York, 2001).

\bibitem{Zanotti}
J.-M. Zanotti, M.-C. Bellissent-Funel, and S.-H. Chen, Relaxational dynamics of supercooled water in porous glass, Phys. Rev. E \textbf{59}, 3084 (1999).

\bibitem{Laage}
D. Laage and W. H. Thompson, Reorientation dynamics of nanoconfined water: Power-law decay, hydrogen-bond jumps, and test of a two-state model, J. Chem. Phys. \textbf{136}, 044513 (2012).

\bibitem{Romero}
S. R.-V. Castrillon, N. Giovambattista, I. A. Aksay, and P. G. Debenedetti, Effect of surface polarity on the structure and dynamics of water in nanoscale confinement, J. Phys. Chem. B \textbf{113}, 1438–1446 (2009).

\bibitem{Trachenko}
K. Trachenko and A. Zaccone, Slow stretched-exponential and fast compressed-exponential relaxation from local event dynamics, J. Phys.: Condens. Matter \textbf{33}, 315101 (2021).




\end{thebibliography}
\end{document}